\documentclass[%
 reprint,
 amsmath,amssymb,
 aps,
twocolumn, pre
]{revtex4-2}

\usepackage{graphicx}
\usepackage{dcolumn}
\usepackage{bm}
\usepackage{hyperref}
\hypersetup{
    colorlinks=true,
    linkcolor=blue,
    filecolor=magenta,      
    urlcolor=blue,
    citecolor=blue,
}

\usepackage[normalem]{ulem}

\begin{document}

\preprint{APS/123-QED}

\title{Molecular communication in one-dimensional channels with active transport and crowding}

\author{Phanindra Dewan}
 \email{phanindrad@iisc.ac.in}
\author{Sumantra Sarkar}%
 \email{sumantra@iisc.ac.in}
\affiliation{%
 Department of Physics,
 Indian Institute of Science, Bengaluru, Karnataka, India - 560012 
}%




\date{\today}

\begin{abstract}
Molecular communication (MC) is a model of information transmission where the signal is transmitted by information-carrying molecules through their physical transport from a transmitter to a receiver through a communication channel. Prior efforts have identified suitable ``information molecules" whose efficacy for signal transmission has been studied extensively in diffusive channels (DC). Although easy to implement, DCs are inefficient for distances longer than tens of nanometers. In contrast, molecular motor-driven nonequilibrium or active transport can drastically increase the range of communication and may permit efficient communication up to tens of micrometers. In this paper, we investigate how active transport influences the efficacy of molecular communication, quantified by the mutual information between transmitted and received signals. We consider two specific scenarios: (a) active transport through relays and (b) active transport through a mixture of active and diffusing particles. In each case, we discuss the efficacy of the communication channel and discuss their potential pitfalls. 
\end{abstract}

\maketitle



\section{Introduction}

Communication systems are designed to transfer information across space and time. Therefore, any communication system must have a transmitter, a receiver, and a channel to transmit information from the transmitter to the receiver~\cite{farsad2016comprehensive}. Modern technologies allow seamless electromagnetic communication in diverse conditions, yet in various situations of importance, electromagnetic wireless communication is unreliable or technologically difficult. Wireless communication is also challenging in extremely small spatial scales, such as in cellular environments, where the ability of cells to communicate and respond to stimuli is vital for many biological processes. Therefore, in such situations, alternative modes of communication are needed. Molecular communication (MC) is an alternative communication model in which information is encoded in processes involving molecules (e.g., concentration or timing data) and transmitted through a molecular transport channel. This paradigm of communication is essential in biological systems that communicate via signaling molecules across many different scales. For example, communication through pheromones in social insects, quorum sensing in bacteria, small molecules, such as Ca$^{2+}$ or cyclic-AMP, diffusion in cell signaling networks, and molecular motor-driven information transfer, such as in neuronal junctions, are all examples of molecular communications~\cite{phillips2012physical, lim2014cell, sarkar2020presence, ngo2020anionic}. While many of these examples occur in 3D, molecular motor-driven communication is an example of 1D molecular communication, which we investigate in this manuscript. Even though it is an idealized system, it provides useful insights into the general principles of MC that can be applied to more realistic and complex scenarios.

MCs require transmitters, receivers, channels, and information carriers, as in any communication system. The encoding used in the communication model determines whether it is a timing channel \cite{eckford2009timing}, an inscribed matter channel \cite{cobo2010bacteria}, or a mass transfer channel \cite{farsad2011quick, eckford2010microchannel}. Although some macroscopic realizations of MCs have been demonstrated \cite{farsad2013tabletop}, most experimental efforts have focused on creating ``information molecules" and the channels through which they are transmitted \cite{gregori2010new,cobo2010bacteria,farsad2016comprehensive}. For example, in kinesin-microtubule~ \cite{hiyama2009biomolecular,hiyama2010biomolecular,hiyama_molecular_2010,enomoto2011design,nitta2006simulating,nitta2010silico}, or myosin-actin-based assays  \cite{byun2009functionalization,ikuta2014tug,steuerwald2014nanoshuttles}, vesicles attached to molecular motors can be thought of as the information molecule, and the filaments can be considered a network of 1D channels. Theoretical and modeling efforts have focused on various aspects of communication engineering, including modeling the receiver, the transmitter, and the channels \cite{moore2009molecular,farsad2011simple,farsad2012mathematical,farsad2014markov}. Because of the presence of memory in MC channels, conventional theoretical treatments have been difficult \cite{farsad2016comprehensive}. However, several important discoveries have been made in understanding the consequences of transport in the so-called diffusion channels \cite{selimkhanov_accurate_2014,hiyama_molecular_2010,aminian_capacity_2015,pierobon_capacity_2013,arifler_capacity_2011,hao_tunable_2013,hoffmann_ikappab-nf-kappab_2002,santos_growth_2007,purvis_encoding_2013,de_ronde_effect_2010,tostevin_mutual_2009,dieterle2020dynamics,cheong_information_2011,tkacik_information_2016,kadloor2012molecular,sarkar2023efficacy,eckford2007nanoscale} and active, nonequilibrium channels 
\cite{sarkar2023efficacy,dieterle2020dynamics}. For example, in \cite{pierobon_capacity_2013}, the authors examined the practical feasibility of MC channels.

Another class of efforts has been targeted at understanding how channel capacity changes due to the mode of transport of the molecules \cite{pierobon_capacity_2013, kadloor2012molecular,gregori2010new,cobo2010bacteria}. At the cellular level, the primary modes of transport are either diffusion or molecular motor-driven active transport~\cite{phillips2012physical}. Diffusion is an equilibrium transport process where molecules are transported through their incessant collision with the surrounding molecules, resulting in unbiased random walks. Active transport, in contrast, is a nonequilibrium transport process which converts a fuel, such as ATP, to mechanical work required for transport~\cite{ramaswamy2010mechanics}. Common examples include animal locomotion and the transport facilitated by molecular motor proteins, which consume ATP to transport a molecule in a biased fashion.

A simple model for understanding active transport is a biased random walk. In a previous work, we investigated whether active transport is always a better option than diffusive transport in MC channels ~\cite{sarkar2023efficacy}. We showed that diffusive transport is a better alternative for short communication channels, which may have led to the evolution of the cellular signaling cascades found in modern cells. For example, in bacteria, where the communication distances are shorter, two-step DCs are predominant, whereas, in eukaryotes, where the communication has to be carried over much larger distances, signals are transmitted through many intermediate steps, where molecules are often transmitted diffusively~\cite{lim2014cell}. Conversely, all long-distance communication in eukaryotes is carried by molecular motors. A well-known example is the transmission of signals through axon~\cite{phillips2012physical,milo2010bionumbers}.

In this paper, we explore the role of active, nonequilibrium transport in MC channels in two manifestations and investigate their consequences in the context of 1D channels. Specifically, we have considered channels where active transport happens (a) through static molecular relays and (b) through a mixture of active and passive particles. Physically, case (a) investigates the variation of MI with the strength of \emph{extrinsic} nonequilibrium drive, and case (b) investigates the same for \emph{intrinsic} drives.  These scenarios have been motivated by experimental observations and prior theoretical works. For example, studies have shown that information transmission in cells might be improved by having relays, in which a molecule can trigger the release of the same type of molecule~\cite{dieterle2020dynamics,strickland2024self}. Relay signaling motifs, which act as an extrinsic drive for signaling molecules in neutrophils, can lead to diffusive information waves~\cite{dieterle2020dynamics}. Motor proteins walking on a microtubule consume ATP and thus have intrinsic drive. In some cases, these motors can bind to specific proteins on the microtubule and drive them, such as in the case of Tau proteins in neurons~\cite{HINRICHS201238559}. Such a situation can give rise to active-passive mixtures along the microtubule. Alternatively, molecular motors such as kinesin can shuttle non-motor microtubule-associated proteins (MAPs) by biasing their diffusion without binding to them~\cite{Farhadi2025.06.28.662138}. Both of these cases are examples of systems having intrinsic drive. Additionally, MC through active transport can be a way of using aspects of communication theory to understand biological signaling networks \cite{nakano2017molecular}.

\section{Methods}

\subsection{Model} 
\begin{figure}[!hbtp]
    \centering
\includegraphics[width=\linewidth]{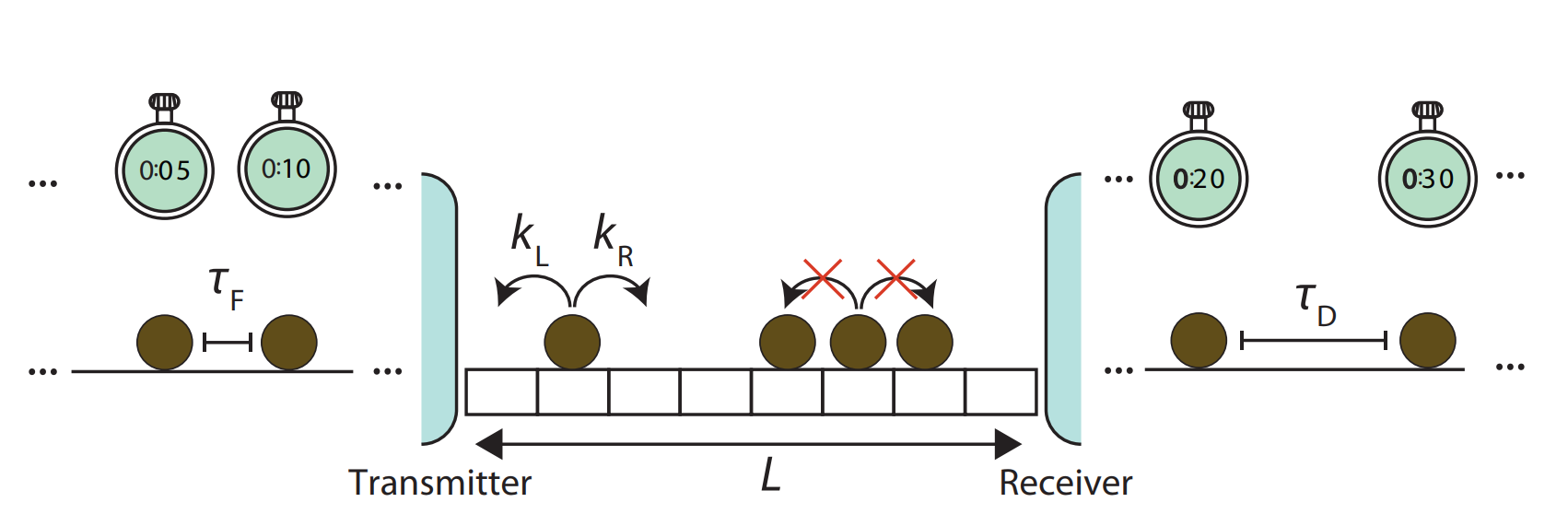}
    \caption{Schematic diagram of the model. A 1D lattice communication channel of length $L$ between the transmitter (left) and the receiver (right). Molecules (brown) are emitted from the transmitter, after which they hop on the lattice with a probability of hopping to the right being $k_R$ and hopping to the left being $k_L$. The molecules have excluded-volume interactions, meaning that no two molecules can occupy the same lattice point at the same time. The firing times $\tau_F$ and the detection times $\tau_D$ are recorded for processing. This schematic has been adopted from Ref.~\cite{sarkar2023efficacy}.}
    \label{fig:fig1}
\end{figure}

We consider a simple one-dimensional (1D) model of molecular communication, where a diffusible molecule stochastically moves from a transmitter to a receiver through a 1D channel. The model and its consequences have been described earlier in Ref.~\cite{sarkar2023efficacy}. Here we summarize the main results for completeness. In this model (Fig.~\ref{fig:fig1}), transport happens on a 1D lattice, where molecules jump towards the receiver (to the right) with rate $k_R$ and towards the transmitter (to the left) with rate $k_L$. The position of a molecule $i$, $x^i$, on the lattice follows the following time evolution:
\begin{equation}
    x^{i}_{t+1} = x^{i}_{t} + \mathcal{X}
\end{equation}
where $\mathcal{X} = 1$ for forward steps with rate $k_R$ and $\mathcal{X} = -1$ for backward steps with rate $k_L$. The molecules are featureless particles that interact via excluded-volume interactions. Hence, only one molecule can occupy a lattice point at a time. For the sake of simplicity, it is assumed that once transmitted, the transmitter cannot reabsorb the molecules, and, once detected, the molecules cannot be emitted back to the lattice by the receiver. These two criteria are accommodated by reflecting boundary conditions at $x = 0$ and absorbing boundary conditions at $x = L$. Although in real biological systems receptors might release signaling molecules back into the bulk after detection, to simplify our analyses, we consider only a specific scenario in which the signaling molecule is absorbed by the receptor. We implement this choice using an absorbing boundary conditions at the detector. This simplification enables us to focus our investigation only on the effect of active transport on the efficacy of information transmission. Our approach is motivated by the classic work on chemoreception by Berg and Purcell \cite{berg1977physics}, who used the same approach to investigate the role of molecular diffusion on the limit of chemoreception, a prime example of MC. As noted in \cite{berg1977physics}, non-absorbing boundary conditions do not change the results qualitatively as long as the absorption rate is high enough. 

 Molecules are emitted or fired from the transmitter at different times, and the intervals between firing times of two consecutive molecules are the firing time intervals and are defined as $\tau_F = t_F^{i+1} - t_F^{i}$. The time interval between two consecutive detection events at the receiver is the detection time interval and is defined as $\tau_D = t_D^{j+1} - t_D^{j}$. $t_F^i$ is the firing time of $i^{th}$ firing event and $t_D^j$ is the detection time of the $j^{th}$ detection event. The detection time for the $j$th molecule is thus, $t_D^j = t^j_F + t^j$, where $t^j$ is the first passage time for a molecule doing a random walk on a lattice. This is, in general, a difficult quantity to obtain, in this case, where molecules are crowded by others.

Although many works on MCs have focused on carrying information in a single molecule, we consider an alternative mode of information transport. We assume that the information is stored in the time interval between firing events, $\tau_F$, which is a realistic assumption in many biological systems~\cite{berg1977physics,BOULWARE2008R769, bettenworth2022frequency} as well as in conventional communication systems, such as the Morse code. For example, it is possible to estimate the concentration of molecules by counting the number of times ligands bind to a receptor. Each binding event may trigger downstream signaling, analogous to a firing event in our model \cite{mehta2012energetic}. Although earlier work uses the firing time and detection times to calculate mutual information \cite{eckford2009timing}, other works on trajectory mutual information show that information transmission in biochemical networks can be calculated between the entire trajectories of the input and output variables \cite{tostevin_mutual_2009, PhysRevResearch.5.013032, PhysRevLett.131.128401}. The trajectory for our event-based system is represented minimally by time intervals rather than absolute times. Additionally, cells can use time interval data for accurately sensing their environment~\cite{berg1977physics,PhysRevLett.103.158101}. In the classic work by Berg and Purcell~\cite{berg1977physics}, the accuracy of concentration estimation depends on the bound and unbound time intervals of ligands to receptors, which are approximated in an average sense. Thus, timing-interval data is the raw data available for cells in sensing their environment. Further, Ref.~\cite{PhysRevLett.103.158101} showed that the full receptor occupied and unoccupied time intervals give an even better accuracy for sensing the concentration. Thus, information can be stored in firing and detection time intervals for our 1D communication channel. The detection time interval, $\tau_D$, is the information received after traveling through the molecular communication channel. Because of the stochastic transport of the information carrier molecules, $\tau_D$ will be statistically different from $\tau_F$. The firing times and the detection times are thus vectors $\mathbf{\tau_F}$ and $\mathbf{\tau_D}$. The detection times do not need to be sorted as in \cite{eckford2007nanoscale}, because in this case the molecules do not overlap, thus preserving their order. Following \cite{tkacik_information_2016},  we estimate the efficacy of information transmission by computing the mutual information between $\tau_F$ and $\tau_D$, defined as~\cite{cover1999elements, holmes2019estimation}: 

\begin{align}
    I(\tau_F,\tau_D) &= \sum_{\tau_F}\sum_{\tau_D}P(\tau_F;\tau_D)\ln \frac{P(\tau_F;\tau_D)}{P(\tau_F)P(\tau_D)}\\
                    &= \sum_{\tau_F}\sum_{\tau_D}P(\tau_F)P(\tau_D|\tau_F) \ln \frac{P(\tau_D|\tau_F)}{P(\tau_D)}
\end{align}
where $P(\tau_F)$ is the firing time distribution, $P(\tau_D)$ is the detection time distribution, and $P(\tau_F;\tau_D)$ is the joint distribution of firing time interval and detection time interval. Further, the information is not stored in the time of release and detection, as calculated in \cite{eckford2009timing}, but is actually stored in the differences between consecutive firing and detection times. This makes the calculation of $I(\tau_F,\tau_D)$ non-trivial. To compare our results across various firing time distributions, we normalized the mutual information by its maximum value, $H(\tau_F)$ :
\begin{equation}
    I = \frac{I(\tau_F;\tau_D)}{H(\tau_F)}
\end{equation}
where $H(\tau_F)$ is the Shannon entropy of the firing time distribution, defined as
\begin{equation}
    H(\tau_F) = - \sum_{\tau_F} P(\tau_F) \ln P(\tau_F)
\end{equation}

Since the information in our model is encoded in the time difference between two molecules, our model is a \emph{timing channel}.  Further, our model accounts for excluded-volume interactions, meaning that no two molecules can occupy the same lattice site at the same time. Thus, crowding has been implicitly included in our model, unlike earlier models that consider independent information molecules. This makes our model more realistic in quasi-1D geometries compared to earlier studies \cite{eckford2010microchannel, farsad2011simple, farsad2014markov, eckford2009timing} when the communication channel might be approximated as a 1D channel where molecules travel from one end to another.

We have carried out stochastic simulations of the 1D model and stored the firing time interval and detection time interval data. In general, estimating MI from samples of two random variables is not an easy problem. However, using methods in \cite{holmes2019estimation}, we can obtain the MI between $\tau_F$ and $\tau_D$~\cite{sarkar2023efficacy}. It is important to note that below a certain threshold for the MI values, of the order of $~10^{-3}$ to $10^{-4}$, MI calculated using this method \cite{holmes2019estimation} does not give a good estimation. Also, the minimum sample sizes for firing and detection time intervals were set to 10000, given that current algorithms for calculating MI are strongly sample-size-dependent \cite{holmes2019estimation, sarkar2023efficacy}.

\subsection{Information transmission in bare channels} 
In Ref.~\cite{sarkar2023efficacy}, we studied the variation of $I$ with the different parameters of the model. The most striking observation was that the variation of $I$ depended only on a dimensionless parameter, called the Péclet number $Pe = vL/D$, where $v = a(k_R - k_L)$ is the drift velocity and $D = a^2(k_R + k_L)/2 $ is the diffusion coefficient of the molecules. Specifically, $I(Pe)$ showed three clear regimes: 
\begin{equation}
    I(Pe) =\begin{cases}
     {Pe}^0 & Pe < Pe_0 = \frac{2v\tau_F}{L} \\
     {Pe}^{-4} & Pe_0 \leq Pe < 1 \\ 
     {Pe}^{-1} & Pe \geq 1
    \end{cases}
\end{equation}
At $Pe = 1$, molecular transport transitions from a diffusion-dominated regime to an advection-dominated regime. Hence, the concurrent transition in $I (Pe)$ at $Pe = 1$ indicates its strong dependence on the transport processes that drive the molecular communication channel. Hence, in this paper, our main goal is to explore information transmission through two modes of molecular transport processes: relays and active transport. 

\subsection{Active transport elements}

\subsubsection{Relays}
In our model, \emph{relays}, as inspired by biological systems \cite{dieterle2020dynamics, strickland2024self}, are special points on the one-dimensional lattice that can alter the hopping rates of the molecules. Once a molecule hops to a relay, say at site $i$, it is always pushed to the next lattice point, $i+1$, towards the receiver, as long as that point is empty. This means $k_R=1$ and $k_L=0$ for a relay, and it is a valid assumption, as recent experiments show that $\text{Ca}^{2+}$ can exhibit relay waves \cite{dieterle2020dynamics}.

\subsubsection{Active Particles}
In a second model, we include mixtures of \emph{active} and \emph{passive} particles, which are defined below. Active particles consume chemical energy and convert it to mechanical motion, which endows them with directional motility~\cite{ramaswamy2010mechanics}. Molecular motors, such as myosin and kinesin, are examples of self-propelled particles. We model an active particle with totally asymmetric hopping rates, which are $k_L = 0$ and $k_R = 1$ in this paper. In contrast, passive particles, such as diffusing molecules, have symmetric hopping rates ($k_L = k_R = 1$) and undergo an unbiased 1D random walk, in contrast to the active particles, which perform a biased 1D random walk. We vary the fraction of active particles, $f_a$, in the mixture and study its impact on the information transmission. The model parameters have been computed for different molecular motors in our earlier work~\cite{sarkar2023efficacy}. Recent work has shown that kinesin motors can push non-motor MAPs along microtubules~\cite{Farhadi2025.06.28.662138}, and $v\approx1$ for kinesin motors~\cite{sarkar2023efficacy}, which is equivalent to $k_L = 0$ and $k_R = 1$ in our model.

\paragraph*{Kymographs}
To visualize the stochastic trajectories of particles on the lattice, we will show kymographs, which show each particle's position along the x-axis and time along the y-axis. The path a particle traces out on the lattice is then plotted in this $x-t$ plane. Kymographs also provide an intuitive picture of the crowding in the channel.

\paragraph*{Cluster size} Passive particles bounded by active particles can form clusters due to the totally asymmetric movement of the latter, which impacts their transport and the transport of the associated information. Hence, it is necessary to identify the clusters and their sizes, which is relatively easy for 1D channels. There, such clusters will essentially be a configuration of the form $APPPP...PA$, with $A$ denoting an active particle and $P$ denoting a passive particle. The detection time interval, $\tau_D$, between two successive particles in a cluster is equal to 1, even though their $\tau_F$ is usually different from 1. The size of a cluster in our model is then the length of the string of 1s in the detection time interval data. We have measured the unconditional cluster-size distribution for different active fractions, as well as the distribution conditioned on $\tau_F \neq 1$. The difference between these distributions is insignificant. Hence, we present the unconditional cluster-size distribution in this paper. 

\section{Results}

\subsection{Channels with a single relay are Markovian communication channels}

\begin{figure}[hbtp]
    \centering
    \includegraphics[width=\linewidth]{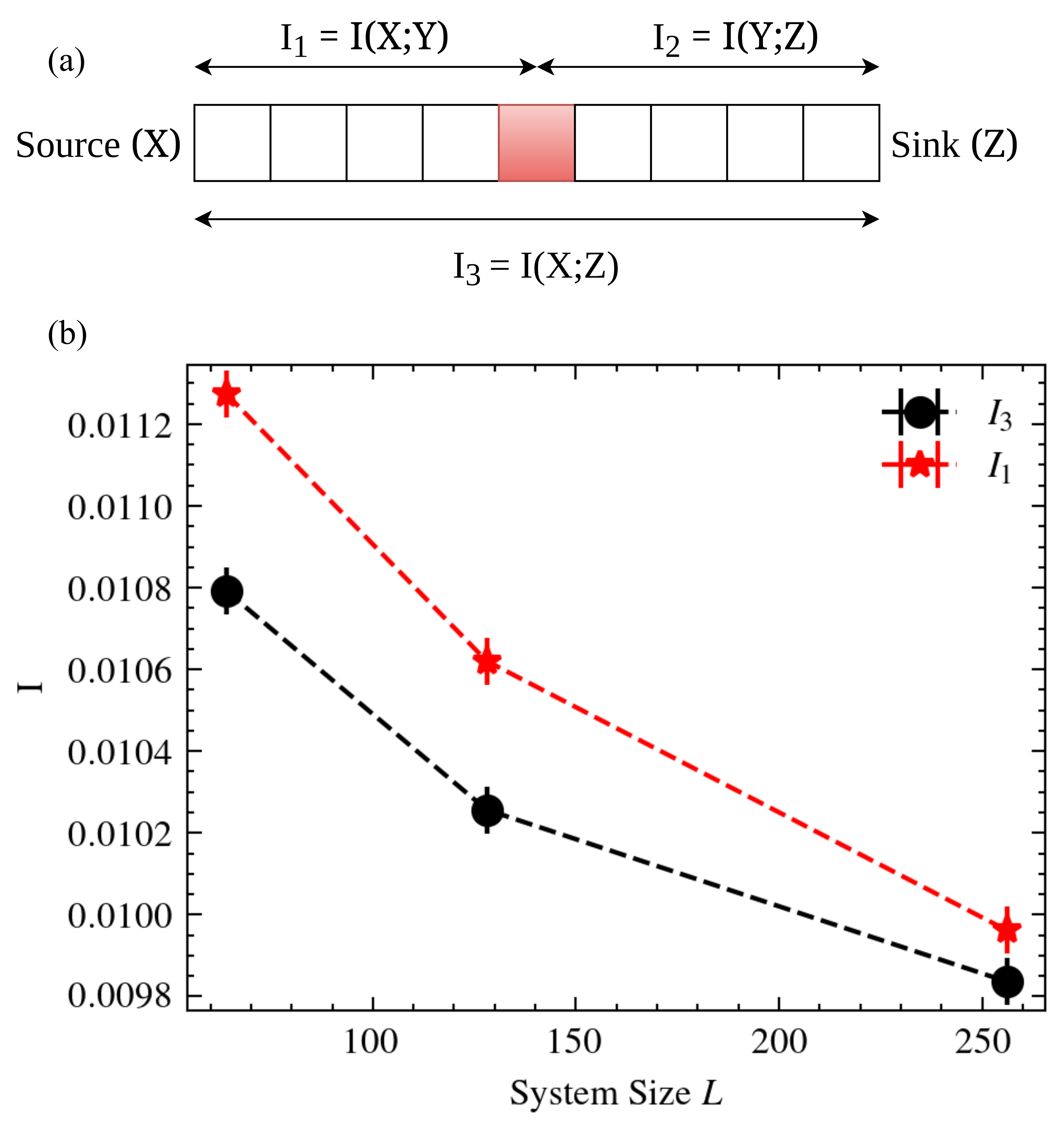}
    \caption{(a) Schematic of the one relay lattice. The red site is the relay. (b) $I_1$ and $I_3$ vs. system size for channels with \emph{one} relay. The relay is kept at the midpoint of the 1D channel.}
    \label{fig:inequality}
\end{figure}

To establish the consistency of our model with the existing models of information transmission, we first describe the variation of mutual information in the presence of just one relay. In this setup, we can characterize the information channel through three different MI for three different transport processes. We denote the MI between the transmitter and the relay as $I_1$, between the relay and the receiver as $I_2$, and between the transmitter and receiver as $I_3$. Because the transport of the molecules is memoryless in the channel, the three processes above form a Markov chain (see Fig. \ref{fig:inequality}a). For a Markov chain $X \rightarrow Y \rightarrow Z$, the \emph{data processing inequality} holds, which states that: $I(X; Y) \geq I(X; Z)$ \cite{cover1999elements}. Here, $I(X; Y)$ is the mutual information between the random variables $X$ and $Y$, and $I(X; Z)$ is the mutual information between $X$ and $Z$. For our model, the communication process boils down to having the following Markov chain: $\tau_F \rightarrow \tau_R \rightarrow \tau_D$. We can denote the MI between emitter and relay as $I(\tau_F;\tau_R) \equiv I_1$ and that between emitter and detector as $I(\tau_F;\tau_D) \equiv I_3$. As shown in Fig.~\ref{fig:inequality}b, $I_1$ is always greater than $I_3$ for all system sizes explored in our 1D model, which is a direct demonstration of the data processing inequality in our model. Hence, we can assume that the model and the MI calculation is consistent with existing models of information transmission.  

\begin{figure*}[!hbtp]
    \centering
    \includegraphics[width=\textwidth]{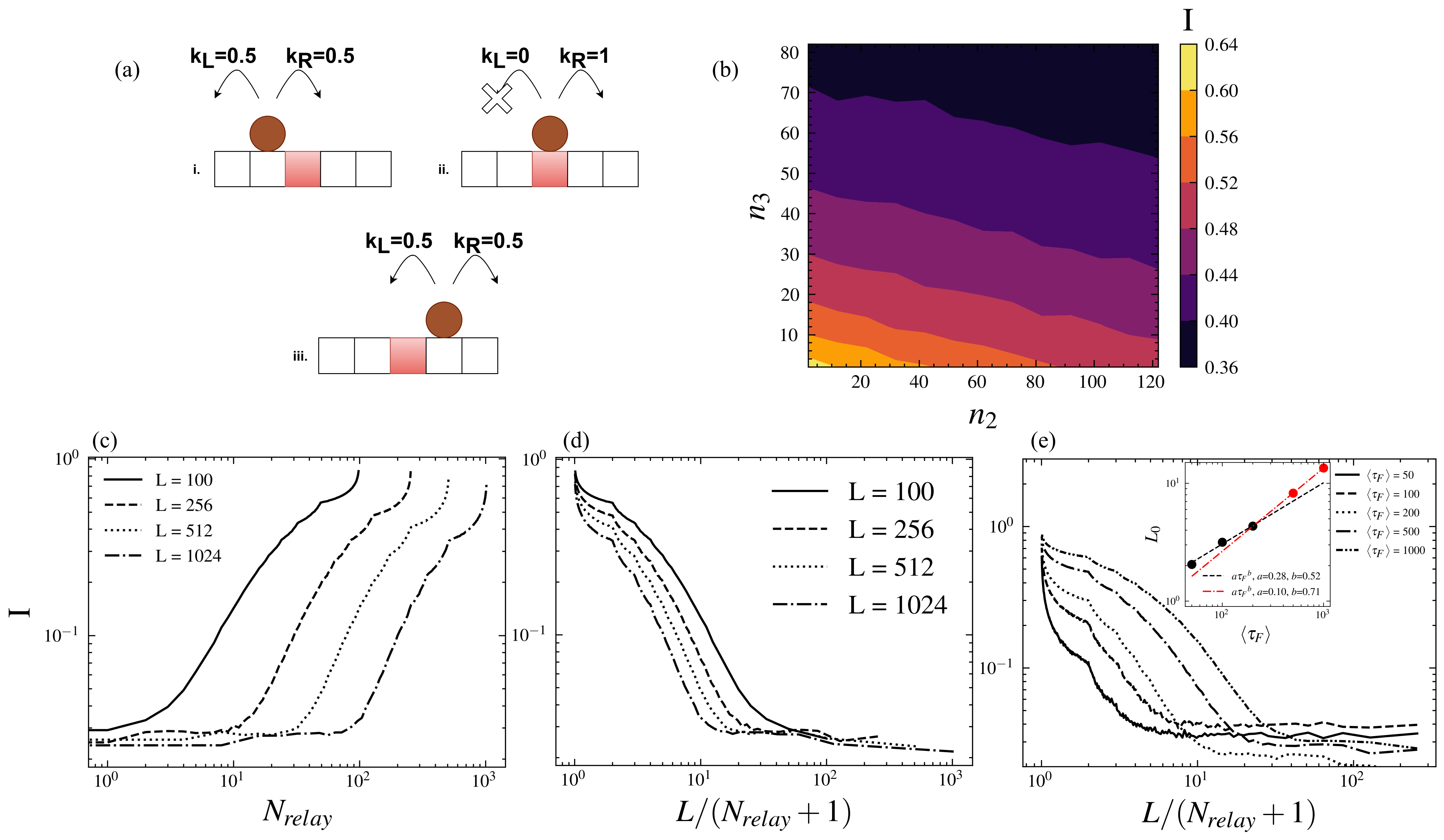}
    \caption{(a) Schematic diagram of relays. The site coloured red is the relay site. (i) A molecule (labeled in brown) approaches the relay site. Its left and right hopping rates are equal, in this case: $k_L=0.5$, $k_R=0.5$. (ii) The molecule reaches the relay site, and its hopping rates change to $k_L=0$ and $k_R=1$. (iii) The molecule is pushed by the relay into the next site, where its hopping rates changes back to $k_L=0.5$ and $k_R=0.5$. (b) A contour plot showing the variation of I, the MI between transmitter-receiver, with $n_2$ and $n_3$, where $n_2$ is the number of DCs of length 2 and $n_3$ is the number of DCs of length 3 in the channel. (c) Variation of MI between source-sink with inter-relay distance for different lengths of the 1D lattice. (d) Variation of MI between source-sink
    with number of relays for different lengths of the 1D lattice. (e) Variation of MI between source-sink with inter-relay distance for different $\left< \tau_F \right>$, with $L=256$. Inset shows variation of threshold length $L_0$ with $\left< \tau_F \right>$. Here, $L_0$ is the threshold inter-relay distance $L/(N_{relay}+1)$ at which $I$ falls below 0.1 in (d).}
    \label{fig:relays}
\end{figure*}

\subsection{Channels with many relays are nontrivial information transporters}
It may be possible to achieve better transmission quality by introducing multiple relays. For example, wireless repeaters can reduce black spots in a wifi network. Also, it has been observed that relay motifs in neutrophils help spread information through diffusive waves~\cite{dieterle2020dynamics}. These observations led us to explore the dependence of MI on the number of relays, $N_{relay}$, in the channel. A schematic of relays in our model is given in \ref{fig:relays}a. This dependence is nontrivial. In the absence of any relays, the channel is purely diffusive and has low efficacy, characterized by a system-size ($L$) dependent MI value. As the number of relays increases,  MI does not increase until an $L$-dependent threshold number of relays, above which it increases non-linearly through repeated exponential rise up to the maximum value of 1 (Fig.~\ref{fig:relays}c). 

Upon closer inspection, we found that mutual information falls exponentially between integer values of inter-relay distance, $L/(N_{relay}+1)$. At each integer value of inter-relay distance, a new exponential fall starts. This is because, at each integer value of inter-relay distance, DCs of longer length start appearing in the channel. For example, if we plot the MI as a function of inter-relay distance ($L/(N_{relay}+1)$), it becomes clear that the end-points of the exponential variations are characterized by integer inter-relay distance, which implies that the distance between all consecutive relays is equal. When the distance is not an integer, then there are spatial variations in the channel length. In both cases, the total information loss across the channel is a combination of the information losses across individual DCs between relays. The exact way in which the information loss across individual DCs adds up to the total information loss is not so clear. Information loss in a DC of length $l$ goes as $l^{-4}$ above a threshold length and as $l^{0}$ below this threshold, $L_0 = \sqrt{2D\langle \tau_F\rangle}$~\cite{sarkar2023efficacy}. 

To understand whether $L_0$ is also the threshold for the nonlinear increase of MI, we computed its variation with inter-relay distance for various values of  $\langle \tau_F\rangle$ shown in Fig.~\ref{fig:relays}e. Indeed, the threshold inter-relay distance increases as  $\langle\tau_F\rangle^{1/2}$ for small values of $\tau_F$, but shows steeper increase at higher values of $\langle \tau_F\rangle$ (Fig.~\ref{fig:relays}e inset). This observation suggests that $L_0$ does govern the threshold inter-relay distance, albeit for a limited range of $\langle \tau_F\rangle$ values. Beyond this insight, unfortunately, we could not find any simple relation between the number of DCs and the information loss. Generalizing the data processing inequality to the multi-relay scenario provided a weak upper bound on the information loss in the entire channel, which was practically unusable. Therefore, we sought to understand the cases where the length of individual DCs was small.

\subsection{Information loss in small diffusive channels follows a stretched exponential law}

\begin{figure}[hbtp]
    \centering
    \includegraphics[width=\linewidth]{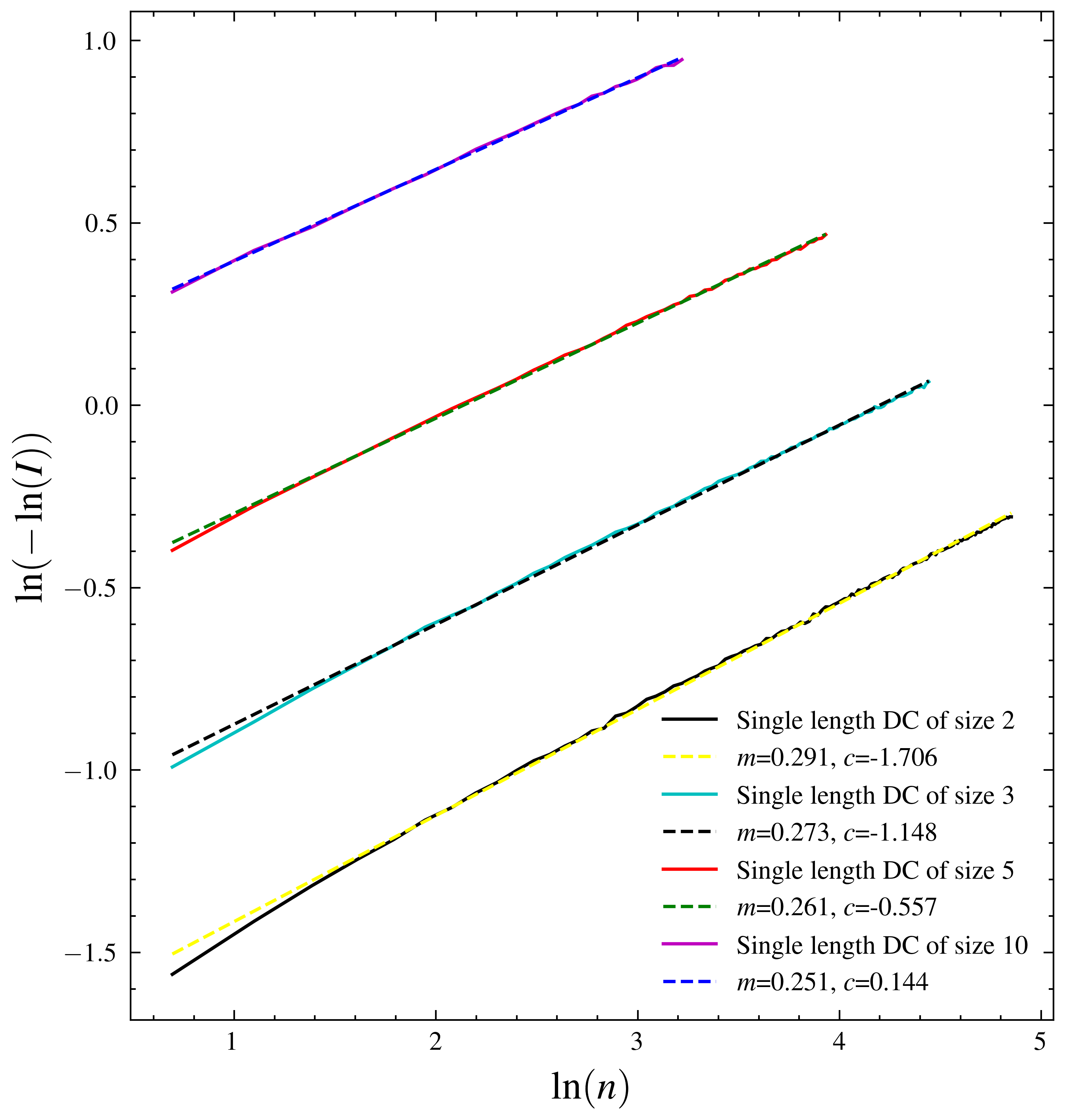}
        \caption{Plot of $\ln(-\ln I)$ vs. $\ln n$, which shows that channels made of single-length DCs of lengths 2, 3, 5, 10 all follow the stretched exponential function given in Eq. \ref{eq: 6}. A straight line is fitted to the simulation data, with the slope, $m$ and intercept $c$ given in the legend.}
    \label{fig:func_DC}
\end{figure}

DCs of varying lengths are the building blocks of the lattice channel with relays. When the inter-relay distance is an integer, the channels are of equal length. For example, when the inter-relay distance changes from 1 to 2, the number of DCs of length 2, say $n_2$, increases from 0 to 64 for a lattice of length 128. Thus, the decrease in MI as inter-relay distance increases is equivalent to the decrease in MI due to an increase in the number of DCs of length 2, $n_2$. Simulation results with channels made of single-length DCs (of lengths 2, 3, 5, and 10) suggested that the MI between the transmitter and receiver made up of DCs of length $l$ is given by a stretched exponential function [Fig. \ref{fig:func_DC}]:
\begin{equation}
\label{eq: 6}
    I(n_l) = e^{-\left(\frac{n_l}{\alpha}\right)^{\mu}} 
\end{equation}
where,
\begin{align*}
        n_l &= \text{number of diffusive channels of length } l\\
        \mu &= \text{stretching exponent}\\
        \alpha &= (\ln(1/I(1)))^{-\frac{1}{\mu}}
\end{align*}

We obtain the above relation by fitting the simulation data to the form suggested above as shown in Fig.~\ref{fig:func_DC}. The best-fit parameters are listed in the table below.

\begin{table}[!hbtp]
    \centering
    \begin{tabular}{|c|c|c|}
    \hline
        $l$ & $\mu$ & $\alpha$ \\
        \hline
        2 & 0.291 & 353.264\\
        3 & 0.273 & 66.585\\
        5 & 0.261 & 8.459\\
        10 & 0.251 & 0.564\\
        \hline
    \end{tabular}
    \caption{$\mu$ and $\alpha$ for different diffusive channel block lengths, $l$.}
    \label{tab:my_label}
\end{table}

\subsection{Differential dependence on block length in mixed relay channels}
The dependence of MI for DCs of a fixed length with no relays was shown before~\cite{sarkar2023efficacy}, but to understand the dependence of MI on $N_{relay}$, we need to understand the dependence of MI for lattice channels with DCs of different lengths, interspersed with relays. This seems to be a difficult task as it is not clear what the functional dependence of MI will be on $n_l$, where $n_l$ is the number of DCs of length $l$. To gain insight, we performed simulations by varying $n_2$ and $n_3$ in mixed channels of length $L = 2n_2 + 3n_3$. As shown in Figure ~\ref{fig:relays}b, the dependence of MI on $n_3$ is much stronger than on $n_2$; roughly, 10 size-3 (total length 30) channels reduce the information by the same amount as 40 size-2 channels (total length 80). This is physically understandable as DCs of length 3 are more likely to decorrelate the molecular transport than DCs of length 2.

\subsection{Increasing active fraction leads to nonlinear increase in MI}

\begin{figure*}
    \centering
    \includegraphics[width = \textwidth]{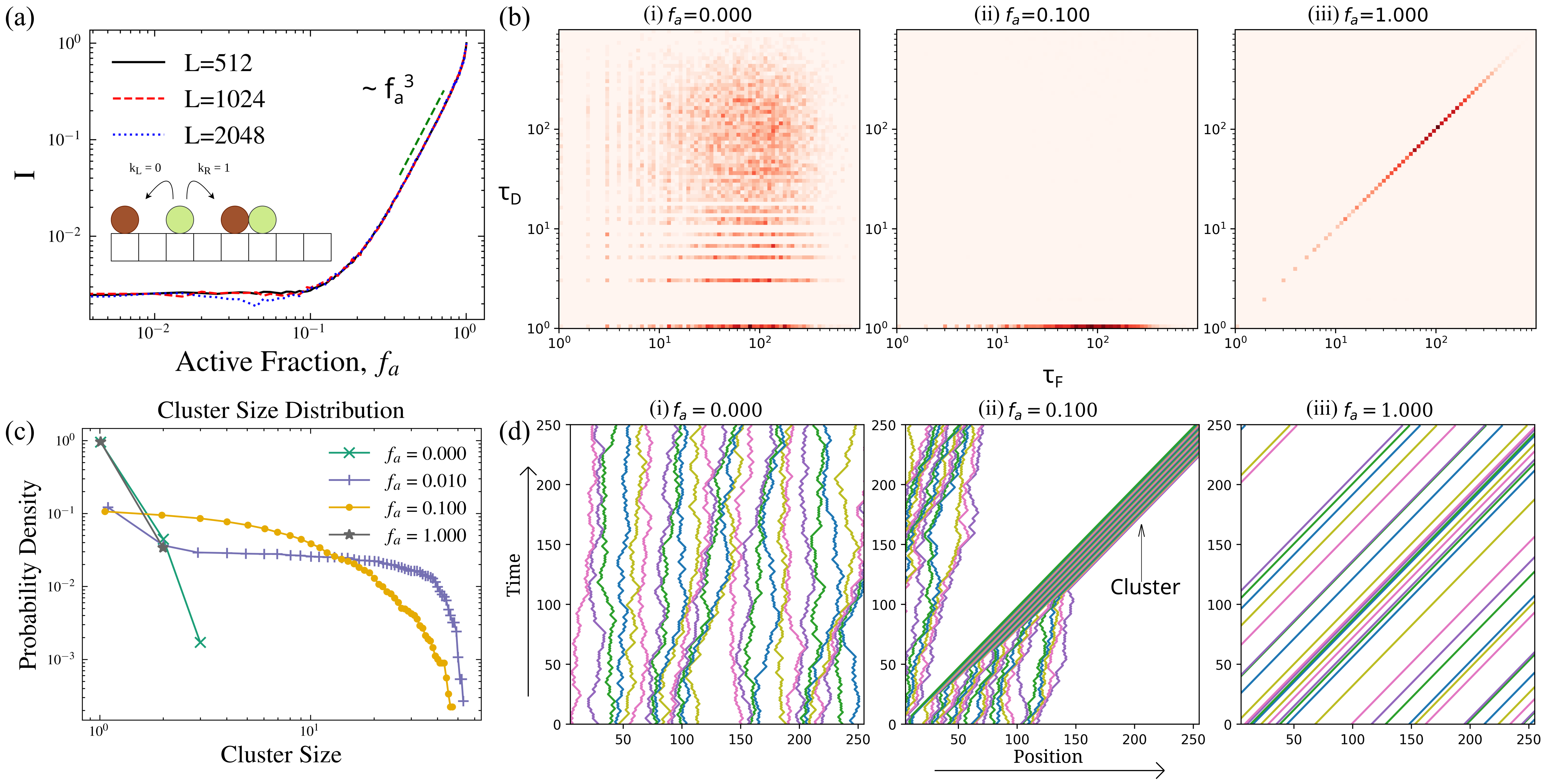}
    \caption{(a) MI vs. active fraction for channels of different lengths, with $k_R=1$ for active particles. The best fit shows that above an active fraction, MI $\sim$ $f_a^3$. The inset shows a schematic showing active particles labeled green, with passive particles labeled brown. (b) The joint probability distribution function, $P(\tau_F, \tau_D)$ for different active fractions, $f_a$. (i) $f_a = 0$, (ii) $f_a = 0.1$, (iii) $f_a = 1$. (c) Cluster size distribution for different active fractions $f_a$. (d) Kymographs from stochastic simulations. Kymographs show molecular trajectories for different active fractions, (i) $f_a=0$, (ii) $f_a=0.1$, and (iii) $f_a=1$, all for $L=256$.}
    \label{fig:active}
\end{figure*}

MI increases non-linearly with active fraction, $f_a$, in a manner (Fig.~\ref{fig:active}a) not unlike the relay channels (Fig.~\ref{fig:relays}c). However, unlike the relay channels, neither the threshold active fraction ($f_a \approx 0.1$ ) nor the general trend depends on the total channel length. The origin of the threshold MI value, which is around $2\times 10^{-3}$ is unclear. Because the algorithm to compute MI has a resolution of $\sim 10^{-3}$~\cite{holmes2019estimation}, the threshold can arise due to the inability of the algorithm to resolve MI variation below this limit. In contrast, the variation of MI above this threshold is well-defined and follows a power law: MI $\sim f_a^{3}$ (Fig.~\ref{fig:active}a).   

\begin{equation}
    I(f_a) =\begin{cases}
     {f_a}^0 & f_a \leq 0.1 \\ 
     {f_a}^{3} & f_a > 0.1
    \end{cases}
\end{equation}

Power law variation implies the presence of a strong correlation in the underlying phenomena. In MC channels, the transport of the molecules determines the variation in MI. Therefore, we suspected that the molecules were moving from the transmitter to the receiver through a collective transport process. Inspection of the joint distribution of $\tau_F$ and $\tau_D$ strengthened this hypothesis (Fig.~\ref{fig:active}b). We found that for $f_a = 0$, the distribution showed no correlation between the variables, and for $f_a = 1$, $\tau_D$ was proportional to $\tau_F$. However, for the threshold value, $f_a = 0.1$, for a wide range of $\tau_F$ values the distribution of $\tau_D$ was centered at $\tau_D = 1$. This observation implied that even though the molecules were fired randomly, they reached the detector one after the other, i.e., in a collective fashion. Indeed, a measurement of the cluster size distribution revealed the presence of large clusters (cluster size of $\sim 20-50$) at small nonzero values of $f_a$, whereas the average cluster size was approximately 1 or 2 for $f_a = 0$ or $1$ (Fig.~\ref{fig:active}c). We found visual support of this observation from the kymographs of the particles, which showed strong clustering for $f_a = 0.1$, but no clustering for $f_a = 0$ or $1$ (Fig.~\ref{fig:active}d). 

\section{Conclusion}
In this paper, we have investigated how the active transport of molecules affects molecular communication in a simple 1D model of molecular communication, proposed in \cite{sarkar2023efficacy}. In this model, we assumed that the information is encoded in the time interval between two consecutive detection events at the receiver, $\tau_D$, which were transmitted from the transmitter with some known time interval distribution, $\tau_F$. We measured the efficacy of the molecular communication channel through the mutual information, $I(\tau_F;\tau_D)$, between $\tau_F$ and $\tau_D$ in two models of active transport.

Our main conclusion is that the crowding of the molecules in the channel is the key driver of the efficacy. Only in the limit of rare firing events ($\tau_F \gg 1$), the active or passive nature of the communication channel becomes important. A striking example of this observation is shown in Fig.~\ref{fig:active}, which shows the transition of the joint probability distribution with the fraction of active particles, $f_a$. Beyond a threshold value of $f_a$, the joint probability distribution at the transition point shows that $\tau_D = 1$ for almost all values of $\tau_F$ [Fig.~\ref{fig:active}b]. Such an extreme distribution arises due to the effect of crowding and the resultant collective transport. At this $f_a$, the configurations are such that the channel gets crowded rapidly, and the particles can only move forward because of their excluded volume interactions. This interaction results in a single-file movement, with particles reaching the receiver immediately after one another. 

The effect of crowding is also important for the relays, where crowding, predominantly present at low values of $\langle \tau_F\rangle$, reduces the mutual information (Fig.~\ref{fig:relays}e). This result is counterintuitive but can be explained through the following reasoning. Intuitively, in DCs, we expect the crowding to reduce the noise arising from diffusive movements. However, the presence of relays and crowding together completely suppresses the diffusive movement. Naively, we expect this effect to increase the MI. However, strong crowding and the relays can make $P(\tau_D)$ a narrowly peaked distribution, which is quite different from the exponential $P(\tau_F)$. Hence, MI can be reduced due to the effect of crowding. It seems that the relays and the active particles have opposing reactions to the crowding of molecules, in addition to their inherent differences in the nature of the nonequilibrium drive. It will be interesting to investigate the combined effects of these special cases. However, that is a formidable task and is beyond the scope of the current manuscript. 

In our model, intrinsic and extrinsic nonequilibrium drives have different effects on molecular communication. The effect of extrinsic drive on MI provided by the relays is system-size dependent (Fig.~\ref{fig:relays}c,d), whereas that of intrinsic drive is not (Fig.~\ref{fig:active}a). The system-size dependence in the presence of relays mirrors the previous result \cite{sarkar2023efficacy}, where, for a system with identical particles, MI depends on system size. In contrast, MI is independent of system size in a mixture of active and passive particles (Fig.~\ref{fig:active}a). Furthermore, in our simulations, MI increases exponentially with the number of relays but as a power law with the fraction of active particles. Hence, in the case of relays, the MI is highly sensitive to the number of relays and may not be a robust mechanism of information transfer for large channels. In contrast, in the active-passive mixture, the changes in MI are far more robust to changes in the fraction of active particles and are insensitive to channel length. This is a remarkable result, and shows that for long channels, active-passive mixtures will fare much better than relays. This may also explain why neuronal signaling is molecular motor-driven or electrochemical wave-driven, but cell signaling in smaller cells occurs through signaling cascades~\cite{lim2014cell}.


It is also worth considering the implications of our work from the perspective of the engineering of MC channels. Active-passive mixtures create a robust transport channel, but maintaining the active particles out of equilibrium is energetically costly. In contrast, although a large number of relays are needed to create a reliable MC channel, the associated energy requirement is far less than the active-passive mixtures, since only the relay site needs to be maintained out of equilibrium, something that can be achieved quite cheaply through a valve. Such consideration is essential given that energetic cost, speed, and accuracy cannot be optimized simultaneously \cite{mehta2012energetic}. Understanding the interplay of these trade-offs will be interesting to explore in the context of molecular communication. 


In conclusion, we sought to investigate the differential effects of active and passive transport in various conditions and found that crowding is the main driver of the efficacy of information transmission. Therefore, to design better molecular communication channels, we should understand the effects of crowding. Crowding is a challenging problem to tackle analytically, as it introduces memory effects. However, in some circumstances, crowding and its effects can be analyzed through analytical techniques \cite{mallick2015exclusion}. Applying these results in the context of molecular communication will be useful and potentially important. 

\section{Code Availability}
The code used can be found in: \url{https://github.com/SSarkarGroup/1d_Mol_Comm_w_Active_Transport}.

\begin{acknowledgments}
 SS would like to thank Axis Bank Centre for Mathematics and Computing, SERB-DST (SERB-22-0223), and IISc for financial support. PD thanks IISc for Ph.D. fellowships.
\end{acknowledgments}


\bibliography{references}

\end{document}